\input{aipcheck}

\documentclass[
    ,final            
  ]
  {aipproc}

\layoutstyle{6x9}

\begin{document}

\title{Photo-assisted shot noise\\ in the fractional quantum Hall regime}


\author{Adeline Cr\'epieux}{
  address={Centre de Physique Th\'eorique, Universit\'e de la M\'editerran\'ee, Case 907, 13288 Marseille, France}
}

\author{Pierre Devillard}{
  address={Centre de Physique Th\'eorique, Universit\'e de Provence, Case 907, 13288 Marseille, France}
}

\author{Thierry Martin}{
  address={Centre de Physique Th\'eorique, Universit\'e de la M\'editerran\'ee, Case 907, 13288 Marseille, France}
}

\begin{abstract}

The effect of an ac perturbation on the shot noise of a fractional quantum Hall fluid is studied at finite temperature. For a normal metal, it is known that the zero-frequency noise derivative exhibits steps as a function of bias voltage. In contrast, at Laughlin fractions, the backscattering noise exhibits evenly spaced singularities, which are reminiscent of tunneling density-of-states singularities for quasiparticles. The spacing is determined by the quasiparticle charge $\nu e$ and the ratio of the dc bias with respect to the drive frequency. Photo-assisted transport can thus be considered as a probe for effective charges of the quantum Hall effect.

\end{abstract}

\maketitle


\section{Introduction}

In mesoscopic systems, the measurement of 
shot noise makes it possible to probe the effective charges
which flow in conductors, and opens the possibility for studying
the role of the statistics in stationary quantum transport experiments. 
This has been illustrated experimentally and theoretically where the interaction between electrons is 
less important\cite{reznikov,kumar_glattli,lesovik,buttiker,beenakker,martin_landauer} or
when it is more relevant\cite{kane_fisher_noise,chamon_freed_wen,saleur,saminadayar,crepieux}.
The present work deals with the study of photo-assisted shot noise 
in a specific one dimensional correlated system: a Hall bar 
in the fractional quantum Hall regime, for which charge transport occurs 
via two counter-propagating chiral edges states. 

\section{Model}

\begin{figure}\label{schema}
  \includegraphics[height=.15\textheight]{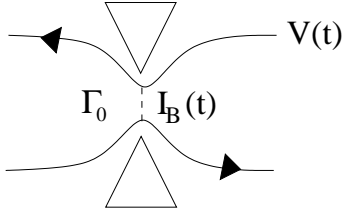}
  \caption{Backscattering between edge states in the presence of a bias voltage modulation $V(t)$.}
\end{figure}

We consider the system depicted on Fig.~\ref{schema} which is described by the Hamiltonian:
\begin{eqnarray}
H=\frac{\hbar \mathrm{v}_F}{4\pi}\sum_{r=R,L}\int ds
(\partial_{s}\phi_r(t))^2+A(t)\Psi_R^\dag(t)\Psi_L(t)+A^*(t)\Psi_L^\dag(t)\Psi_R(t)~.
\label{hamiltonian}
\end{eqnarray}

The bosonic fields $\phi_{R(L)}$, which describe the right and left moving chiral excitations along the edge states, are related to the fermionic fields $\Psi_{R(L)}$ through:
\begin{eqnarray}
\Psi_{R(L)}(t)&=&\frac{F_{R(L)}}{\sqrt{2\pi a}}e^{i\sqrt{\nu}\phi_{R(L)}(t)}~,
\end{eqnarray}

where $F_{R(L)}$ is a Klein factor, $a$, the short-distance cutoff and $\nu$, the filling factor which characterize the charge $e^*=\nu e$ of the backscattered quasiparticles. The hopping amplitude between the edge states has a time-dependence due to the applied voltage $V(t)=V_0+V_1\cos(\omega t)$:
\begin{eqnarray}\label{A}
A(t)=\Gamma_0\sum_{n=-\infty}^{+\infty}J_n\left(\frac{\omega_1}{\omega}\right)e^{i(\omega_0+n\omega)t}~,
\end{eqnarray}

where we have made an expansion in term of an infinite sum of Bessel functions, which is a signature of photo-assisted processes\cite{buttiker_landauer_tunneling}. It is important to notice that the frequencies $\omega_0$ and $\omega_1$ which appear in Eq.~(\ref{A}) are related to the filling factor $\nu$:
\begin{eqnarray}
\omega_0\equiv \nu e V_0/\hbar~,\;\;\;\;\;\;\;\;\;\;\;\;\;\;\omega_1\equiv \nu e V_1/\hbar~.
\end{eqnarray}

where $\nu=1/(2m+1)$ with $m$ integer.

\section{Photo-assisted shot noise}

The symmetrized backscattering current noise correlator is expressed with the help of the Keldysh contour:
\begin{eqnarray}
S(t,t')&=&{1\over 2}\langle I_B(t)I_B(t')\rangle+{1\over 2}\langle I_B(t')
I_B(t)\rangle -\langle I_B(t)\rangle\langle I_B(t')\rangle\nonumber\\
&=&\frac{1}{2}\sum_{\eta=\pm 1}\langle T_K\{I_B(t^\eta)I_B(t'^{-\eta})e^{-i\int_K dt_1H_B(t_1)}\}\rangle
~,
\label{noise_definition}
\end{eqnarray}

where $H_B$ is the sum of the second and the third terms in Eq.~(\ref{hamiltonian}), and:
\begin{eqnarray}
I_B(t)=\frac{i\nu e}{\hbar}A(t)\Psi_R^\dagger(t)\Psi_L(t)-h.c.
\end{eqnarray}

 We are interested in the Poissonian limit only, so 
in the weak backscattering case, one collects the second order contribution in 
the tunnel barrier amplitude $A(t)$, and the product of the average backscattering 
currents can be dropped. The meaning of the Poissonian limit 
is that quasiparticles which tunnel from one edge to another
do so in an independent manner. Yet by doing so they can absorb
or emit $n$ ``photon'' quanta of $\omega$ ($n$ integer). The main purpose of this work is to analyze the double Fourier transform of the noise $S(\Omega_1,\Omega_2)\propto\int dt\int dt' S(t,t')\exp(i\Omega_1t+i\Omega_2t')$ when both 
frequencies $\Omega_1$ and $\Omega_2$ are set to zero. Indeed, the presence 
of the AC perturbation mimics a finite frequency noise measurement. At zero temperature, the shot noise exhibit divergences at each integer value of the ratio $\omega_0/\omega$\cite{crepieux2}. These divergences are not physical since they appear in a range of frequencies where the perturbative calculation turn out to be no more valid. For this reason, we have performed finite temperature calculations which prevent divergences in the backscattering current and shot noise. At finite temperature, the shot noise reads:
\begin{eqnarray}
S(0,0)
&=&\frac{(e^*)^2\Gamma_0^2}{2\pi^2a^2{\overline{\Gamma}}(2\nu)}
\left(\frac{a}{\mathrm{v}_F}\right)^{2\nu}
\left(\frac{2\pi}{\beta}\right)^{2\nu-1}\nonumber\\
&&\times\sum_{n=-\infty}^{+\infty}J_n^2\left(\frac{ \omega_1}{\omega}\right)
\cosh\left(\frac{(\omega_0+n\omega)\beta}{2}\right)\left|{\overline{ \Gamma}}\left(\nu+i\frac{(\omega_0+n\omega)\beta}{2\pi}\right)\right|^2~,
\end{eqnarray}

where $\overline{\Gamma}$ is the Gamma function and $\beta=1/k_BT$.

\begin{figure}\label{nu_un}
  \includegraphics[height=.35\textheight]{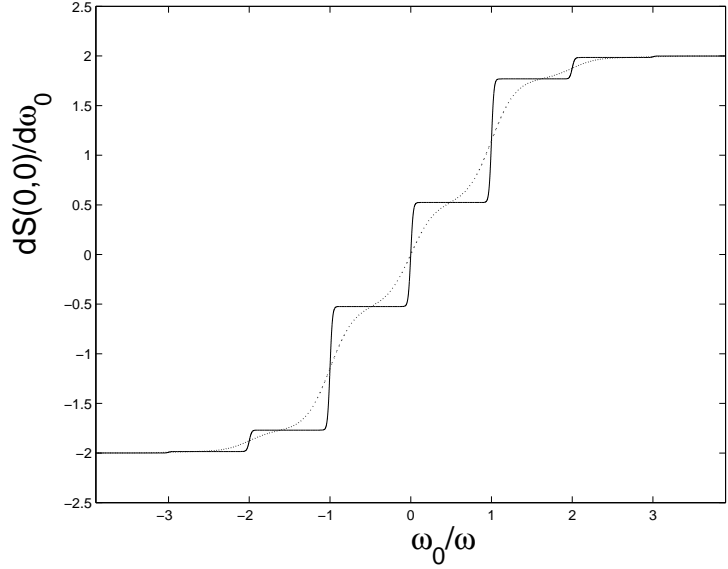}
  \caption{Noise derivative for a normal metal at different temperatures: $k_BT/\hbar\omega=0.01$ (solid line) and $k_BT/\hbar\omega=0.1$ (dashed line). We take $\omega_1/\omega=eV_1/\hbar\omega=3/2$.}
\end{figure}

\begin{figure}\label{nu_un_tiers}
  \includegraphics[height=.35\textheight]{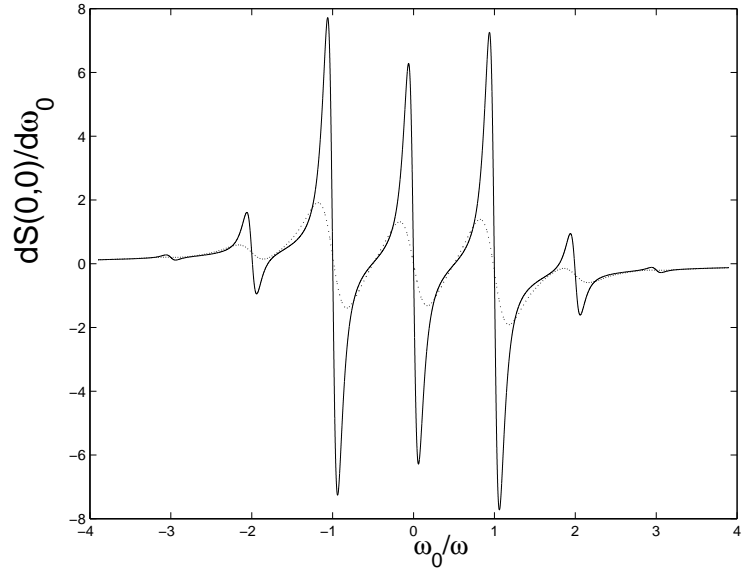}
  \caption{Noise derivative in the fractional quantum Hall regime at a filling factor $\nu=1/3$ for $k_BT/\hbar\omega=0.05$ (solid line) and $k_BT/\hbar\omega=0.15$ (dotted line). We take $\omega_1/\omega=\nu eV_1/\hbar\omega=3/2$.}
\end{figure}

\section{Discussion}

We have first test the validity of our result by setting $\nu=1$, the value which corresponds to non-interacting system (normal metal). The derivative of the shot noise according to the bias voltage exhibits staircase behavior as shown on Fig.~\ref{nu_un}. Steps occur every time  $\omega_0$ is an integer multiple of the ac frequency. This is in complete agreement with the results obtained be Lesovik and Levitov for a Fermi liquid\cite{lesovik2}. When the temperature increases, the steps are rounded.

For non-integer value of the filling factor ($\nu=1/3$, for example), the shot noise derivative exhibits evenly spaced singularities, which are reminiscent of the tunneling density of states singularities for Laughlin quasiparticles. The spacing is determined by the quasiparticle charge $\nu e$ and the ratio of the bias voltage with respect to the ac frequency, and the amplitude is governed by temperature (see Fig.~\ref{nu_un_tiers}). Photo-assisted transport can thus be considered as a probe for effective charges at such filling factors, and could be used in the study of more complicated fractions of the quantum Hall effect.


\end{document}